\documentclass[preprint,showpacs,preprintnumbers,amsmath,amssymb]{revtex4}
\usepackage{amssymb}

\usepackage{graphicx}
\usepackage{dcolumn}
\usepackage{bm}

\begin{document}

\title{Experimental investigation of the non-Markovian dynamics of classical and quantum correlations}

\author{Jin-Shi Xu}
 \affiliation{Key Laboratory of Quantum Information,
  University of Science and Technology
  of China, CAS, Hefei, 230026, People's Republic of China}
 \author{Chuan-Feng Li$\footnote{email: cfli@ustc.edu.cn}$}
\affiliation{Key Laboratory of Quantum Information,
  University of Science and Technology
  of China, CAS, Hefei, 230026, People's Republic of China}
 \author{Cheng-Jie Zhang}
\affiliation{Key Laboratory of Quantum Information, University of
Science and Technology of China, CAS, Hefei, 230026, People's
Republic of China}
 \author{Xiao-Ye Xu}
\affiliation{Key Laboratory of Quantum Information, University of
Science and Technology of China, CAS, Hefei, 230026, People's
Republic of China}
 \author{Yong-Sheng Zhang}
\affiliation{Key Laboratory of Quantum Information, University of
Science and Technology of China, CAS, Hefei, 230026, People's
Republic of China}
\author{Guang-Can Guo}
\affiliation{Key Laboratory of Quantum Information, University of
Science and Technology of China, CAS, Hefei, 230026, People's
Republic of China}
\date{\today }
\begin{abstract}
We experimentally investigate the dynamics of classical and quantum
correlations of a Bell diagonal state in a non-Markovian dephasing
environment. The sudden transition from classical to quantum
decoherence regime is observed during the dynamics of such kind of
Bell diagonal state. Due to the refocusing effect of the overall
relative phase, the quantum correlation revives from near zero and
then decays again in the subsequent evolution. However, the
non-Markovian effect is too weak to revive the classical
correlation, which remains constant in the same evolution range.
With the implementation of an optical $\sigma_{x}$ operation, the
sudden transition from quantum to classical revival regime is
obtained and correlation echoes are formed. Our method can be used
to control the revival time of correlations, which would be
important in quantum memory.
\end{abstract}

\maketitle

\section{Introduction}

Quantum entanglement, as a kind of distinctive quantum correlation
without the classical counterpart, is widely recognized as the
crucial resource of quantum communication and computation
\cite{Nielsen00}. However, there are also other nonclassical
correlations that can even exist in separated quantum states
\cite{Ollivier01,Henderson01}. The finding that nonclassical
correlations other than quantum entanglement may provide the speedup
in the deterministic quantum computation with one pure qubit (DQC1)
protocol \cite{Datta08,Lanyon08} has greatly motivated the related
study.

One of the essential issues is to distinguish different kinds of
correlations in quantum systems. In the classical information
theory, correlation is perfectly characterized by the Shannon
entropy, which is represented by the classical mutual information
\cite{Shannon48}. When it comes to quantum systems, the Shannon
entropy is replaced by the von Neumann entropy and the total
correlation in a bipartite system is characterized by the quantum
mutual information \cite{Groisman05}. For bipartite systems, the
quantumness of correlation can be quantified by the quantum discord
which represents the difference between classical information theory
and quantum information theory \cite{Ollivier01}. A related method
concerning classical correlation is proposed and it is based on the
maximal information one can extract with a one-sided local
measurement \cite{Henderson01}. It has been demonstrated that almost
all quantum states contains non-vanishing quantum discord that is
composed by the quantum entanglement and non-entanglement quantum
correlation \cite{Ferraro10}.

There are many investigations of quantum correlation measured by
quantum discord in different kinds of physical systems, including
the spin chains \cite{Sarandy09,Werlang10,Chen10,Werlang102},
Jaynes-Cummings systems \cite{Cole10}, spin-boson systems
\cite{Ge01}, optical systems \cite{Xu10}, quantum dots
\cite{Fanchini-Castelano10} and NMR systems \cite{Soares10}.
Recently, this kind of measurement is also extended to continuous
variable systems \cite{Giorda10,Adesso10,Vasile10} and even the
quantum biology systems \cite{Bradler10}. Quantum discord has been
found to be useful in quantum information processing and quantum
information theory. The non-entanglement quantum discord may provide
the quantum advantage in the DQC1 quantum computation protocol
\cite{Datta08,Lanyon08}. The vanishing of quantum discord implies
completely positive dynamics maps \cite{Shabani09}. Quantum discord
also plays important roles in many basic physical problems, for
example the Maxwell's demon \cite{Zurek03}, quantum phase transition
\cite{Sarandy09,Werlang10,Chen10,Werlang102} and the relative effect
\cite{Datta09,WangDeng10}.

The quantification of quantum discord is based on the one-sided
measurement on a bipartite system, which is usually asymmetric. We
would get different results by choosing different subsystem to be
measured. A symmetrical method with two-sided measurement over both
partitions of a bipartite system is proposed to quantify the
classical correlation which is represented by the maximal classical
mutual information \cite{Terhal02,DiVincenzo04}. There are also
other approaches proposed to distinguish classical and quantum
correlations. Inspired by the work consumption when extracting
information from a heat bath, a thermodynamics approach is used to
defined quantum correlation in quantum systems
\cite{Oppenheim02,Horodecki05}. In particularly, the difference
between the extractable information with the closed local operation
and classical communication and the total information is defined as
the quantum information deficit, which is used to quantify quantum
correlation \cite{Horodecki05}. Because classical states are
measured without disturbance, classical and quantum correlations are
also characterized by the measurement-induced disturbance
\cite{Luo08}. Recently, by employing the relative entropy as a
distance measure of correlations, the method proposed by Modi {\it
et al.} \cite{Modi10} provides an unified view of quantum and
classical correlations. Different from quantum discord based on
bipartite mutual information, this method can be extended to
quantified different kinds of correlations in multipartite systems
of arbitrary dimensions. Generally, all these measurements of
classical and quantum correlations mentioned above are not equal to
each other due to the different definitions. However, they are
consistent in the case of distinguishing classical and quantum
correlations in Bell diagonal states \cite{Xu10}.

Another interesting subject is to investigate the dynamics of
different kinds of correlations in noisy environments. For one side,
the inevitable interaction between a quantum system and its
environment would destroy correlations in the system and leads to
the reduction of useful resource. The knowledge of the dynamic
behavior of correlations will help us to design suitable protocols
to protect correlations under processing. For the other side,
stimulated by the discovery of distinctive dynamic behavior of
entanglement sudden death \cite{Yu09}, that is, entanglement
disappear completely in a finite evolution time, the investigation
of the unusual dynamic behavior of classical and quantum
correlations has caused great interests. Usually, the dissipative
correlation evolution is essentially dependent on the types of
noises that act on the system. Markovian noises would cause the
irreversible decay of system information into the environment. It
has been shown that quantum discord is more robust than entanglement
under Markovian noises \cite{Werlang09}, in the sense that
entanglement may suffer from sudden death, whereas quantum
correlation decays exponentially. The sudden change in behavior in
the decay rates of classical and quantum correlations has been
predicted in different kinds of Markovian environments
\cite{Maziero09} and has been experimentally verified in an optical
system \cite{Xu10}. The decoherence-free evolution of quantum and
classical correlations under certain Markovian noise and the sudden
transition from classical to quantum decoherence regime are also
observed \cite{Xu10,Mazzola10}. The Markovian dynamics of classical
and quantum correlations is also experimentally investigated in a
nuclear magnetic resonance quadrupolar system \cite{Soares10}. When
it is extended to the non-Markovian environment with memory effect,
the feedback information from the environment to the interested
system may greatly affect the dynamic behavior of correlations.
Quantum discord under non-Markovian noises is shown to be
instantaneously vanished compared to the completely disappearance of
entanglement in a finite time interval \cite{Wang10,Fanchini10}. The
revival of quantum discord without the revival of entanglement is
also demonstrated in non-Markovian noises \cite{Werlang10}.

In this paper, based on the equal footing method with the relative
entropy as a distance measurement of correlations, we experimentally
investigate the dynamics of classical and quantum correlations of a
two-photon Bell diagonal state in a one-sided non-Markovian
environment, which is simulated by a Fabry-Perot cavity followed by
quartz plates. At the beginning of evolution, the sudden transition
from classical to quantum decoherence regime is shown
\cite{Xu10,Mazzola10}. Due to the refocusing effect of the relative
phase in the non-Markovian environment, the quantum correlation
revives from a near zero area and then decays again in the
subsequent evolution. However, the non-Markovian effect is too weak
to revive the classical correlation, which remains constant at the
same evolution range. We then perform a $\sigma_x$ operation on the
photon under decoherence and the sudden transition from quantum to
classical revival regime is obtained, in which correlation echoes
are formed.

The paper is organized as follows. The measure of correlations using
relative entropy and the theoretical description of the experiment
are given in section II. The experimental setup and results are
presented in section III. We then give a discussion and conclusion
in section IV.

\section{Relative entropy of correlations and Theoretical description of the experiment}

\subsection{Relative Entropy of Correlations}

The magnitude of a specified property in a quantum system can be
quantified by the distance from the interested state to the closest
state without that desired property \cite{Modi10}. For example,
quantum entanglement can be characterized by the relative entropy of
entanglement ($REE$) \cite{Vedral97}, which is described as the
minimal distance measured with relative entropy between the state
$\rho$ and a separated state $\sigma$ and is expressed as
$REE=\underset{\sigma\in D}{\min}S(\rho||\sigma)$.
$S(\rho||\sigma)=-\mbox{tr}(\rho\log_{2}\sigma)-S(\rho)$ and
$S(\rho)=-\mbox{tr}(\rho\log_{2}\rho)=-\sum_{j}p_{j}\log_{2}p_{j}$
is the von Neumann entropy ($p_{j}$ represent the eigenvalues of
$\rho$). D is the set of separable states. As a result, if all the
distances are measured with relative entropy, different kinds of
correlations in a quantum system can be measured on an equal
footing. The quantum correlation is then defined as the minimal
distance between $\rho$ and a classical state $\chi$, which is
expressed as $\mathcal{Q}=\underset{\chi\in C}{\min}S(\rho||\chi)$
\cite{Modi10}. $C$ represents the set of classical states. Whereas
the classical correlation is defined as the minimal distance between
$\chi$ and a product state $\pi$ and is expressed as
$\mathcal{C}=\underset{\pi\in P}{\min}S(\chi||\pi)$ \cite{Modi10},
where $P$ is the set of product states.

It has been demonstrated that the calculations of quantum and
classical correlations in a quantum system ($\rho$) can be further
simplified as \cite{Modi10}
\begin{eqnarray}
\mathcal{Q}&=&S(\chi_{\rho})-S(\rho), \label{quantum} \\
\mathcal{C}&=&S(\pi_{\chi_{\rho}})-S(\chi_{\rho}), \label{classical}
\end{eqnarray}
where $\chi_{\rho}$ represents the closest classical state of $\rho$
and $\pi_{\chi_{\rho}}$ is the corresponding reduced state of
$\chi_{\rho}$ in the product form which is the closest product state
of $\chi_{\rho}$.

The total mutual information of $\rho$ can be calculated as
\begin{equation}
\mathcal{I}=S(\pi_{\rho})-S(\rho),  \label{total}
\end{equation}
which represents the minimal distance between $\rho$ and its reduced
state in the product form $\pi_{\rho}$ \cite{Modi10}. For bipartite
systems, $\mathcal{I}$ is equal to the quantum mutual information.

Generally, it is difficult to find the closest classical state
$\chi_{\rho}$. However, for the Bell diagonal state
$\rho=\sum_{i=1}^{4}\lambda_{i}|\Phi_{i}\rangle\langle\Phi_{i}|$
where $\lambda_{i}$ are the non-increasing eigenvalues and
$|\Phi_{i}\rangle$ represent the four Bell states, the analytic
expression of $\chi_{\rho}$ is found and
$\chi_{\rho}=\frac{\lambda_{1}+\lambda_{2}}{2}(|\Phi_{1}\rangle\langle\Phi_{1}|+|\Phi_{2}\rangle\langle\Phi_{2}|)+
\frac{\lambda_{3}+\lambda_{4}}{2}(|\Phi_{3}\rangle\langle\Phi_{3}|+|\Phi_{4}\rangle\langle\Phi_{4}|)$
\cite{Modi10}. As a result, for the Bell diagonal state with the
eigenvalues $\{\lambda_{1},\lambda_{2},\lambda_{3},\lambda_{4}\}$,
the eigenvalues of the closest classical state is
$\{\frac{1}{2}(\lambda_{1}+\lambda_{2}),\frac{1}{2}(\lambda_{1}+\lambda_{2}),
\frac{1}{2}(\lambda_{3}+\lambda_{4}),\frac{1}{2}(\lambda_{3}+\lambda_{4})\}$.
We can therefore calculate the quantum correlation $\mathcal{Q}$
according to equation (\ref{quantum}). The product states of $\rho$
and $\chi_{\rho}$ are both identical and equal to the normalized
identity $\openone/4$. As a result, the classical correlation and
total mutual information can be calculated according to equations
(\ref{classical}) and (\ref{total}) respectively and
$\mathcal{I}=\mathcal{Q}+\mathcal{C}$. The analytic solution of the
relative entropy of entanglement for Bell diagonal states is given
by \cite{Vedral97}
\begin{equation}
REE=1+\lambda_{1}\log_{2}\lambda_{1}+(1-\lambda_{1})\log_{2}(1-\lambda_{1}),
\label{entanglement1}
\end{equation}
if $\lambda_{1}\geq1/2$, whereas
\begin{equation}
REE=0, \label{entanglement2}
\end{equation}
if $\lambda_{1}\in[0,1/2]$.

As a result, with the knowledge of $\rho$, one can compute different
kinds of correlations.

\subsection{Theoretical description of the experiment}

The pure dephasing environment is a kind of uniquely quantum noise,
which causes randomness between the relative phases of information
carries. In optical systems, the coupling between the photon
polarization states (information carriers) and photon frequency
(noise freedoms) in a birefringent environment leads to the
dephasing with a trace over frequency \cite{Berglund00}.

Consider a maximally entangled polarization state of two photons
with the form of
$|\Psi\rangle=\frac{1}{2}(|HH\rangle_{a,b}+|HV\rangle_{a,b}+|VH\rangle_{a,b}-|VV\rangle_{a,b})$,
where $|H\rangle$ and $|V\rangle$ represent horizontal and vertical
polarization states, respectively. The subscripts $a$ and $b$ denote
different paths of these two photons. When the entangled state
passes through dephasing environments, which are simulated by quartz
plates with the optic axes set to be horizontal, the final
polarization state for a certain single frequency can be written as
\begin{align}
|\Psi'\rangle
&=\frac{1}{2}(|HH\rangle_{a,b}+e^{i\phi_{b}}|HV\rangle_{a,b}+e^{i\phi_{a}}|VH\rangle_{a,b} \nonumber \\
&\quad-e^{i(\phi_{a}+\phi_{b})}|VV\rangle_{a,b}),
\end{align}
where $\phi_{a}=L_{a}\Delta n\omega_{a}/c$ and $\phi_{b}=L_{b}\Delta
n\omega_{b}/c$. $L_{a}$ ($\omega_{a}$) and $L_{b}$ ($\omega_{b}$)
represent the thickness of quartz plates (the frequency of the
photon) in paths $a$ and $b$, respectively. $\Delta n$ is the
difference between the indices of refraction of ordinary and
extraordinary light, and $c$ represents the vacuum velocity of the
photon. By tracing over all the frequency degrees of freedom, two
decoherence parameters $\kappa_{a}=\int
g(\omega_{a})e^{i\phi_{a}}\mathrm{d}\omega_{a}$ and $\kappa_{b}=\int
f(\omega_{b})e^{i\phi_{b}}\mathrm{d}\omega_{b}$ would impose on the
reduced polarization density matrix, where $g(\omega_{a})$ and
$f(\omega_{b})$ represent the frequency distributions of the photon
in paths $a$ and $b$ and they are normalized as $\int
g(\omega_{a})\mathrm{d}\omega_{a}=1$ and $\int
f(\omega_{b})\mathrm{d}\omega_{b}=1$, respectively. The final
density matrix in the canonical basis
$\{|HH\rangle,|HV\rangle,|VH\rangle,|VV\rangle\}$ becomes
\cite{Xu102}
\begin{equation}
\rho=\frac{1}{4}\left(
\begin{array}{cccc}
1 & \kappa_{b}^{\ast} & \kappa_{a}^{\ast} & -\kappa_{a}^{\ast}\kappa_{b}^{\ast} \\
\kappa_{b} & 1& \kappa_{a}^{\ast}\kappa_{b} & -\kappa_{a}^{\ast} \\
\kappa_{a}& \kappa_{a}\kappa_{b}^{\ast}& 1&-\kappa_{b}^{\ast} \\
-\kappa_{a}\kappa_{b} & -\kappa_{a}& -\kappa_{b}&1%
\end{array}%
\right),\label{density:mix}
\end{equation}
where $\kappa_{a}^{\ast}$ ($\kappa_{b}^{\ast}$) corresponds to the
complex conjugate of $\kappa_{a}$ ($\kappa_{b}$). This final evolved
state can be transformed into a Bell diagonal form with local
unitary operations. For a special case with $\kappa_{a}$ and
$\kappa_{b}$ both setting to be real, the four eigenvalues of $\rho$
are given by
$\{\frac{1}{4}(1+\kappa_{a})(1+\kappa_{b}),\frac{1}{4}(1-\kappa_{a})(1+\kappa_{b}),
\frac{1}{4}(1+\kappa_{a})(1-\kappa_{b}),\frac{1}{4}(1-\kappa_{a})(1-\kappa_{b})\}$.
Because $0\leq\kappa_{a}\leq1$ and $0\leq\kappa_{b}\leq1$, the
maximal eigenvalue is $\frac{1}{4}(1+\kappa_{a})(1+\kappa_{b})$ and
the minimal eigenvalue is $\frac{1}{4}(1-\kappa_{a})(1-\kappa_{b})$.
Whereas the identification of the second maximal eigenvalue is
dependent on the relative magnitudes of $\kappa_{a}$ and
$\kappa_{b}$. If we set $\kappa_{a}$ to be a fixed value (with a
fixed thickness of $L_{a}$) representing the decoherence parameter
in preparing the initial mixed state and $\kappa_{b}$ to be the
decoherence parameter in the evolution ranging from 1 to 0, the
second maximal eigenvalue is
$\frac{1}{4}(1-\kappa_{a})(1+\kappa_{b})$ when
$\kappa_{a}\leq\kappa_{b}$, whereas the second maximal eigenvalue is
given by $\frac{1}{4}(1+\kappa_{a})(1-\kappa_{b})$ when
$\kappa_{a}>\kappa_{b}$. Therefore, the four non-increasing
eigenvalues of the closest classical state $\chi_{\rho}$ are
$\{\frac{1}{4}(1+\kappa_{b}),\frac{1}{4}(1+\kappa_{b}),\frac{1}{4}(1-\kappa_{b}),\frac{1}{4}(1-\kappa_{b})\}$
when $\kappa_{a}\leq\kappa_{b}$. In the case of
$\kappa_{a}>\kappa_{b}$, the four eigenvalues become
$\{\frac{1}{4}(1+\kappa_{a}),\frac{1}{4}(1+\kappa_{a}),\frac{1}{4}(1-\kappa_{a}),\frac{1}{4}(1-\kappa_{a})\}$,
which are all fixed values. As a result, the quantum correlation is
calculated as
\begin{equation}
\mathcal{Q} = \left\{
 \begin{array}{cc}
      \frac{1}{2}(1+\kappa_{a})\log_{2}(1+\kappa_{a})+\frac{1}{2}(1-\kappa_{a})\log_{2}(1-\kappa_{a}),  & \text{if }~\kappa_{a}\leq\kappa_{b}, \\
      \frac{1}{2}(1+\kappa_{b})\log_{2}(1+\kappa_{b})+\frac{1}{2}(1-\kappa_{b})\log_{2}(1-\kappa_{b}), &  \text{if }~\kappa_{a}>\kappa_{b}. \\
 \end{array}
\right.
\end{equation}
and the classical correlation is expressed by
\begin{equation}
\mathcal{C} = \left\{
 \begin{array}{cc}
     \frac{1}{2}(1+\kappa_{b})\log_{2}(1+\kappa_{b})+\frac{1}{2}(1-\kappa_{b})\log_{2}(1-\kappa_{b}),  & \text{if }~\kappa_{a}\leq\kappa_{b}, \\
      \frac{1}{2}(1+\kappa_{a})\log_{2}(1+\kappa_{a})+\frac{1}{2}(1-\kappa_{a})\log_{2}(1-\kappa_{a}), &  \text{if }~\kappa_{a}>\kappa_{b}. \\
 \end{array}
\right.
\end{equation}

We can find that the quantum correlation and classical correlation
remain constant when $\kappa_{a}\leq\kappa_{b}$ and
$\kappa_{a}>\kappa_{b}$, respectively (for fixed $\kappa_{a}$). They
overlap at the point of $\kappa_{a}=\kappa_{b}$. As a result, the
state of equation (\ref{density:mix}) with $\kappa_{b}=1$ represents
the kind of initial states with the property of exhibiting the
sudden transition from classical to quantum decoherence regime
\cite{Xu10,Mazzola10}.

Generally, the frequency spectrum of the photon is peaked at a
central value $\omega_{0}$ with a finite width $\sigma$, for example
the Gaussian function like frequency distribution
$f(\omega_{b})=(2/\sqrt{\pi}\sigma)\exp[-4(\omega_{b}-\omega_{0})^2/\sigma^{2}]$.
The decoherence parameter $\kappa_{b}$ is therefore calculated as
$\kappa_{b}=\exp[-(L_{b}\Delta n/c)^{2}\sigma^{2}/16+i(L_{b}\Delta
n/c)\omega_{0}]$ and it decays exponentially, which leads to the
Markovian limited dynamics of correlations. However, if the
frequency distribution of the photon in mode $b$ becomes discrete,
such as the combination of finite $N$ Gaussian frequency
distributions
$f(\omega_{b})=\sum_{j=1}^{N}A_{j}(2/\sqrt{\pi}\sigma_{j})\exp[-4(\omega_{b}-\omega_{j})^{2}/\sigma_{j}^{2}]$
where $A_{j}$ are the relative amplitude for each Gaussian function
distribution with the central frequencies $\omega_{j}$ and frequency
widths $\sigma_{j}$. In this case, the decoherence parameter is
calculated as $\kappa_{b}=\sum_{j=1}^{N}A_{j}\exp[-(L_{b}\Delta
n/c)^{2}\sigma_{j}^{2}/16+i(L_{b}\Delta n/c)\omega_{j}]$. During the
dephasing process, the overall relative phase may refocus and the
non-Markovian effect occurs, which leads to the revival of
$\kappa_{b}$. $\kappa_{b}$ may be larger than $\kappa_{a}$ again and
it would give rise to the revival of correlations. In experiment,
the discrete frequency distribution can be realized by passing the
photon in mode $b$ through a Fabry-Perot cavity, which behaves as an
optical resonator \cite{Xu102}. Wavelengths for which the cavity
optical thickness is equal to an integer multiple of half
wavelengths are resonant in the cavity and transmitted. Other
wavelengths within the reflective band of the Fabry-Perot cavity are
reflected.

By controlling the non-Markovian effect, we can get the revival of
$\mathcal{Q}$ without the revival of $\mathcal{C}$ (the maximal
revival value of $\kappa_{b}$ is less than $\kappa_{a}$), and the
case that both $\mathcal{Q}$ and $\mathcal{C}$ get revival (the
maximal revival value of $\kappa_{b}$ larger than $\kappa_{a}$).
Actually, the maximal revival of $\mathcal{Q}$ and $\mathcal{C}$ can
be realized by completely compensating the decoherence in the pure
dephasing environment. If we exchange the polarizations of
$|H\rangle$ and $|V\rangle$ of the photon in mode $b$ during the
dynamics, the randomness of the relative phase caused in the
previous evolution time is compensated by the same subsequent
evolution time \cite{Berglund00}. For the state $|\Psi'\rangle$, it
becomes $|\Psi'\rangle
=\frac{1}{2}e^{i\phi_{b}}(|HV\rangle_{a,b}+|HH\rangle_{a,b}+e^{i\phi_{a}}|VV\rangle_{a,b}
-e^{i\phi_{a}}|VH\rangle_{a,b})$. As a result, the effect of
decoherence in mode $b$ reduces to an unobservable global phase and
the final state is changed to the initial form with the exchanging
of $|H\rangle$ and $|V\rangle$ in mode $b$ again. In our experiment
described below, we employed such spin echo like technology
\cite{Hahn50} to obtained the maximal revival of both $\mathcal{Q}$
and $\mathcal{C}$.

\section{Experimental setup and results}

\begin{figure}[tbph]
\begin{center}
\includegraphics [width= 3in]{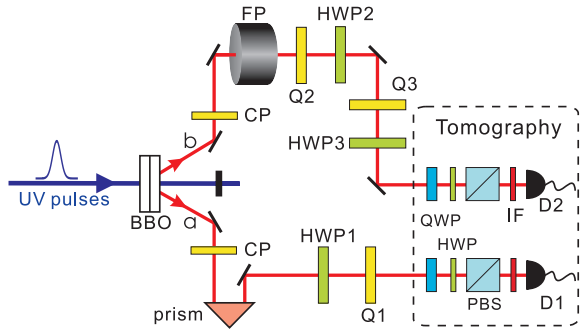}
\end{center}
\caption{(Color online). Experimental setup. Maximally entangled
polarization photon pairs, created from the spontaneous parametric
down conversion process by pumping the two-crystal geometry type I
BBO with ultraviolet (UV) pulses, are emitted into paths $a$ and
$b$. These two photons pass through quartz plates (CP) to compensate
the birefringence in BBO. A half-wave plate (HWP1) and quartz plates
(Q1) in path $a$ are used to prepared the required mixed state. The
Fabry-Perot (FP) cavity together with quartz plates (Q2 and Q3) and
two half-wave plates (HWP2 and HWP3) in path $b$ is employed to
monitor the dephasing coupling. After passing through quarter-wave
plates (QWP), half-wave plates (HWP) and polarization beam splitters
(PBS) which allow to tomographically reconstruct the density matrix,
both photons are then registered by single-photon detectors (D1 and
D2) equipped with 3 nm interference filters (IF).} \label{fig:setup}
\end{figure}

Figure \ref{fig:setup} shows the experimental setup to investigate
the correlation dynamics in the non-Markovian environment. The
second harmonic ultraviolet (UV) pulses are frequency doubled from a
mode-locked Ti:sapphire laser with the center wavelength mode locked
to 0.78 $\mu$m (with a 130 fs pulse width and a 76 MHz repetition
rate). These UV pulses are then focused into two thin, identically
cut type-I $\beta$-barium borate (BBO) crystals with their optic
axes aligned perpendicularly to each other \cite{Kwiat99}.
Degenerate photon pairs with wavelengthes centred at 0.78 $\mu$m,
created from the spontaneous parametric down conversion process, are
emitted into paths $a$ and $b$. By compensating the birefringence
effect in the BBO crystals with quartz plates (CP) in both paths,
the prepared maximally entangled state
($1/\sqrt{2}(|HH\rangle_{a,b}+|VV\rangle_{a,b})$) can has a high
purity \cite{Xu06}.

A half-wave plate (HWP1) with the optic axis set at $22.5^{\circ}$,
which changes $|H\rangle$ into $1/\sqrt{2}|H+V\rangle$ and
$|V\rangle$ into $1/\sqrt{2}|H-V\rangle$, is used to transfer the
maximally entangled state into the exact form of $|\Psi\rangle$.
Quartz plates (Q1) with the optic axis set to be horizontal are used
to dephase the photon in the path $a$. The frequency distribution
$g(\omega_{a})$ is considered as a continuous Gaussian function,
which is defined by the interference filter with a 3 nm full width
at half maximum (FWHM).

The frequency distribution of the photon in path $b$
($f(\omega_{b})$) becomes discrete after it passes through the
Fabry-Perot (FP) cavity, which is a 0.2 mm thick quartz glass with
coating films (reflectivity 90\% at wavelengths around 780 nm) on
both sides \cite{Xu102}. Quartz plates (Q2 and Q3) with the optic
axes set to be horizontal are used to introduce the dephasing
effect. A half-wave plate (HWP2) with the optic axis set to
$45^{\circ}$ acting as a $\sigma_{x}$ operation can exchange
$|H\rangle$ and $|V\rangle$. Another half-wave plate (HWP3) with the
same setting as HWP2 after Q3 is used to change the final form of
the output state.

The density matrix of the final state is reconstructed by the
quantum state tomography process \cite{James01}, where the 16
different coincidence measurement bases are set by quarter-wave
plates (QWP), half-wave plates (HWP) and polarization beam splitters
(PBS). Both photons are then detected by single-photon detectors
equipped with 3 nm interference filters to give coincidence counts.

\begin{figure}
\centering
\begin{minipage}[c]{0.5\textwidth}
\centering
\includegraphics[width=2.5in]{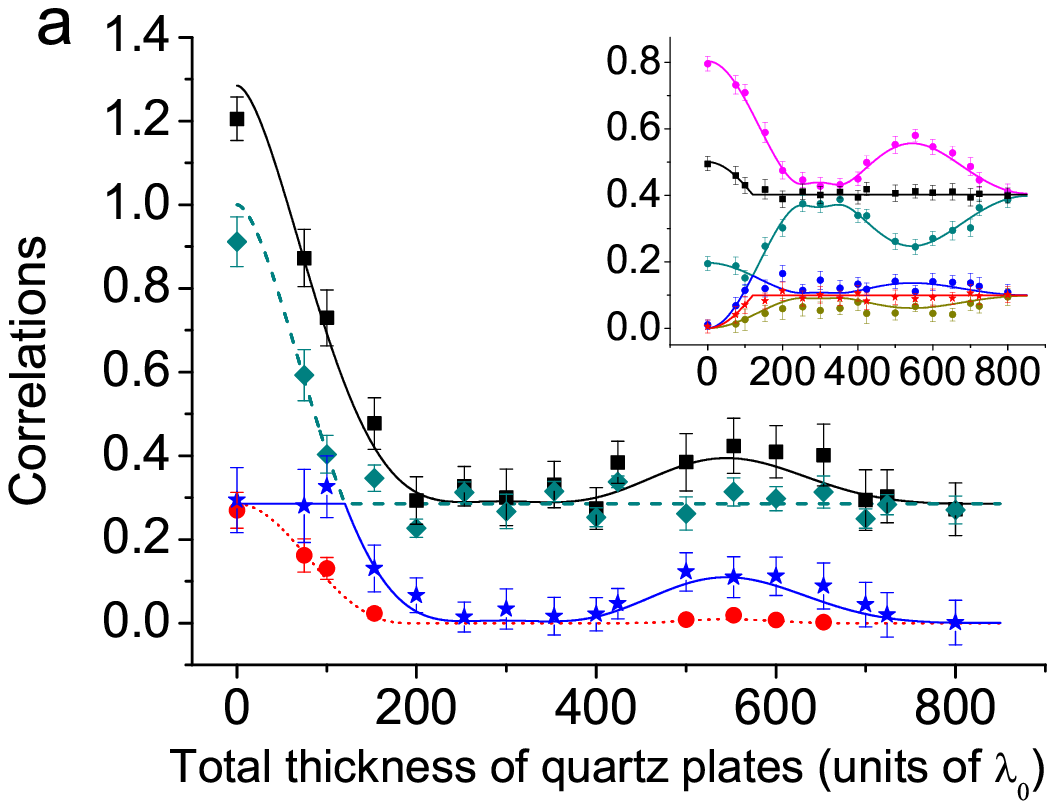}
\end{minipage}\\
\begin{minipage}[c]{0.5\textwidth}
\centering
\includegraphics[width=2.5in]{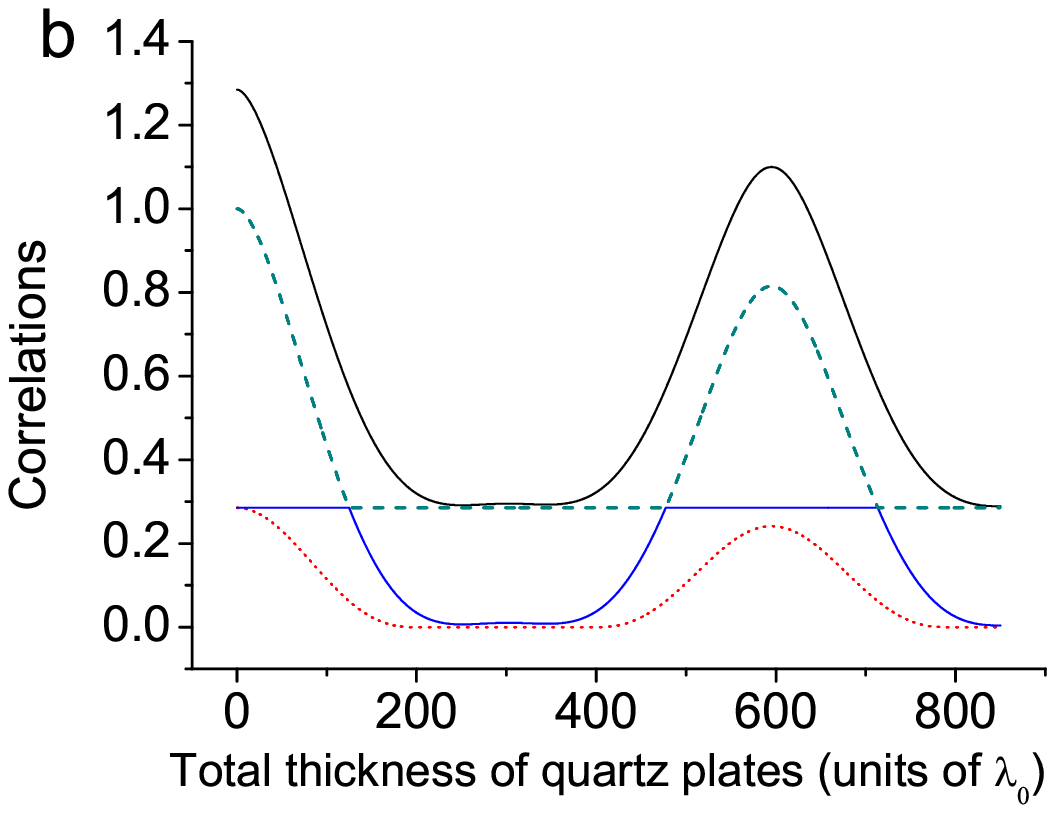}
\end{minipage}\\
\caption{(Color online). ({\bf a}) Experimental results of the
non-Markovian dynamics of correlations. Black squares, dark cyan
diamonds, blue stars and red dots represent the experimental results
of $\mathcal{I}$, $\mathcal{C}$, $\mathcal{Q}$ and $REE$ with the
black solid line, dark cyan dashed line, blue solid line and red
dotted line representing the corresponding theoretical prediction,
respectively. The inset (the $x$ axes represent the total thickness
of Q2) represents the corresponding evolution of eigenvalues of
$\rho$ and $\chi_{\rho}$. The magenta, dark cyan, blue and dark
yellow dots represent the experimental results of $\lambda_{1}$,
$\lambda_{2}$, $\lambda_{3}$ and $\lambda_{4}$ with the magenta,
dark cyan, blue and dark yellow solid lines representing the
corresponding theoretical predictions, respectively. Black squares
and red stars represent the experimental results of
$\frac{1}{2}(\lambda_{1}+\lambda_{2})$ and
$\frac{1}{2}(\lambda_{3}+\lambda_{4})$ with the black and red solid
lines representing the corresponding theoretical predictions,
respectively. ({\bf b}) Theoretical predictions of the correlation
dynamics with the FWHM of $f(\omega_{b})$ identically fitting to 0.2
nm. The black solid line, dark cyan dashed line, blue solid line and
red dotted line represent $\mathcal{I}$, $\mathcal{C}$,
$\mathcal{Q}$ and $REE$, respectively. $\lambda_{0}=0.78$ $\mu$m.}
\label{fig:experiment}
\end{figure}

In our experiment, the final output state is a Bell diagonal state
with the form of equation (\ref{density:mix}). According to the
equal footing method \cite{Modi10}, experimental results of quantum
correlation ($\mathcal{Q}$), classical correlation ($\mathcal{C}$)
and total correlation ($\mathcal{I}$) are deduced from equations
(\ref{quantum}), (\ref{classical}) and (\ref{total}) with
$S(\pi_{\chi_{\rho}})=S(\pi_{\rho})$, respectively. The relative
entropy of entanglement ($REE$) can be calculated from the maximal
eigenvalue of $\rho$ according to equations (\ref{entanglement1})
and (\ref{entanglement2}) \cite{Vedral97}. As a result, all kinds of
correlations can be deduced from the density matrix of $\rho$.

Fig. \ref{fig:experiment} displays the dynamics of correlations, as
a function of the thickness of Q2 which is represented by the
corresponding retardation $x$ ($L=x/\Delta n$). The thickness of Q1
represented by the retardation is set to be $117\lambda_{0}$
($\lambda_{0}$=0.78 $\mu$m is the central wavelength of the photon)
to prepare the initial mixed state and $\kappa_{a}$ is equal to
about 0.607. The HWP2 and HWP3 are not used ({\it i. e.}, without
$\sigma_{x}$ operation) and Q3 is set to 0, in which the phenomenon
of entanglement collapse and revival occurs \cite{Xu102}. We find
that there exhibits a sudden transition from classical to quantum
decoherence area \cite{Xu10,Mazzola10}. At the beginning of the
evolution, the quantum correlation $\mathcal{Q}$ (blue stars)
remains constant and then decays exponentially after the thickness
of about $120\lambda_{0}$. Due to the refocusing effect of the
relative phase, $\mathcal{Q}$ revives from near zero at the
thickness of about $440\lambda_{0}$ and reaches the maximum of 0.11
at about $540\lambda_{0}$. With further increasing Q2, $\mathcal{Q}$
decays monotonically again. The classical correlation $\mathcal{C}$
(dark cyan diamonds) behaviors quite differently. It decays
exponentially at the beginning and then remains constant all the
time after the thickness of $120\lambda_{0}$ despite the
non-Markovian effect. $\mathcal{Q}$ and $\mathcal{C}$ overlap at the
thickness of $120\lambda_{0}$, in which the sudden change in
behavior in their decay rates are observed \cite{Maziero09,Xu10}.
The evolution of relative entropy of entanglement $REE$ (red dots)
is also shown, which suffers from sudden death \cite{Yu09} at the
thickness of about $189\lambda_{0}$ and is consistent with our
previous results \cite{Xu102}. If we continue to increase Q2, $REE$
also revives to its maximally value at the thickness of about
$540\lambda_{0}$ (the value is relative small in the figure). The
evolution of total correlation $\mathcal{I}$ (black squares) first
decays exponentially and then revives just as that of $\mathcal{Q}$.
The black solid line, dark cyan dashed line, blue solid line and red
dotted line represent the theoretical predictions of $\mathcal{I}$,
$\mathcal{C}$, $\mathcal{Q}$ and $REE$. The inset displays the
corresponding dynamics of eigenvalues of $\rho$ and $\chi_{\rho}$.
The magenta, dark cyan, blue and dark yellow dots represent the
experimental results of $\lambda_{1}$, $\lambda_{2}$, $\lambda_{3}$
and $\lambda_{4}$ with the magenta, dark cyan, blue and dark yellow
solid lines representing the corresponding theoretical predictions,
respectively. Whereas the black squares and red stars represent the
experimental results of $\frac{1}{2}(\lambda_{1}+\lambda_{2})$ and
$\frac{1}{2}(\lambda_{3}+\lambda_{4})$ with the black and red solid
lines representing the corresponding theoretical predictions. It can
be seen clearly that the sudden transition of classical and quantum
decoherence occurs at the point when the switch in the second
maximal eigenvalue $\lambda_{2}$ occurs \cite{Mazzola10} and it is
consistent with our previous theoretical prediction. At the period
with the relative phase refocusing, the four eigenvalues
$\lambda_{i}$ behave correspondingly, {\it i. e.}, $\lambda_{1}$
($\lambda_{3}$) increases and $\lambda_{2}$ ($\lambda_{4}$)
decreases. However, $\frac{1}{2}(\lambda_{1}+\lambda_{2})$ and
$\frac{1}{2}(\lambda_{3}+\lambda_{4})$ remain constant and there is
not revival of $\mathcal{C}$ according to its definition. The errors
of the experimental results are estimated using the method proposed
in ref. \cite{James01}, which is mainly due to the random
fluctuation of each measured coincidence counts (the errors from the
uncertainties in aligning the wave plates is relatively small). In
this approach, the errors are directly deduced from the fluctuation
of the corresponding eigenvalues. The error bars of the relative
entropy of entanglement involve only the maximal eigenvalue of
$\rho$, which are relative small in the figure.

In our experiment, the frequency distribution $f(\omega_{b})$ is
treated as three Gauss-like wave packets centered at 778.853,
780.160 and 781.459 nm with the relative probabilities of 0.37, 0.44
and 0.19, respectively \cite{Xu102}. The FWHM of these wave packets
are identically considered as 0.85 nm and we obtain good fittings.
The maximal revival value of $\kappa_{b}$ is about 0.385, which is
smaller than $\kappa_{a}$. As a result, the classical correlation
remains constant after the sudden transition point in the case of
Fig. \ref{fig:experiment}a, which is immune from the non-Markovian
effect and is consistent with our previous analysis in the
theoretical part. With the narrower FWHM of $f(\omega_{b})$, the
maximal revival value of $\kappa_{b}$ can be larger than
$\kappa_{a}$ and the switch in the second maximal eigenvalue would
occur again, in which the refocusing effect would be strong enough
to revive the classical correlation. Fig. \ref{fig:experiment}b
shows the theoretical prediction of the correlation dynamics with
the FWHM of $f(\omega_{b})$ identically fitting to 0.2 nm (the
maximal revival value of $\kappa_{b}$ is about 0.944). We can see
that $\mathcal{I}$ (black solid line), $\mathcal{C}$ (dark cyan
dashed line), $\mathcal{Q}$ ( blue solid line) and $REE$ (red dotted
line) are all revived. The sudden transition from quantum to
classical revival regime is obtained at the thickness of about
$477\lambda_{0}$ (the revival value of $\kappa_{b}$ is equal to
$\kappa_{a}$).

\begin{figure}
\begin{minipage}[c]{0.5\textwidth}
\centering
\includegraphics[width=2.5in]{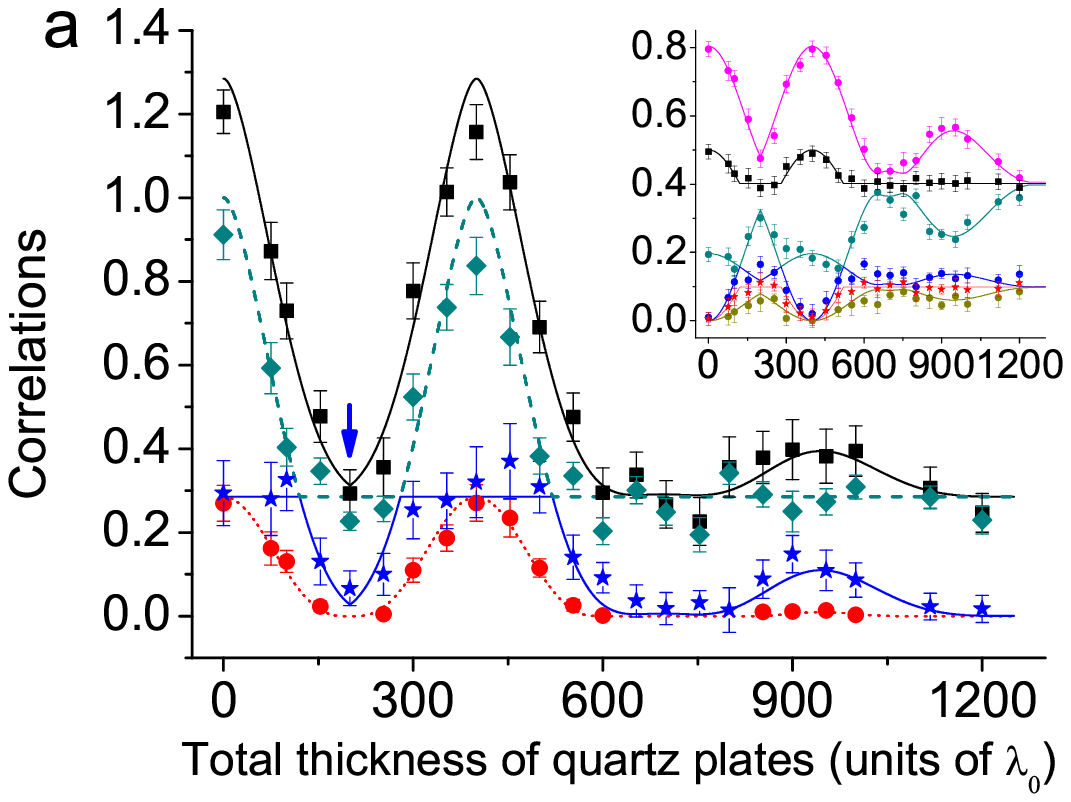}
\end{minipage}\\
\begin{minipage}[c]{0.5\textwidth}
\centering
\includegraphics[width=2.5in]{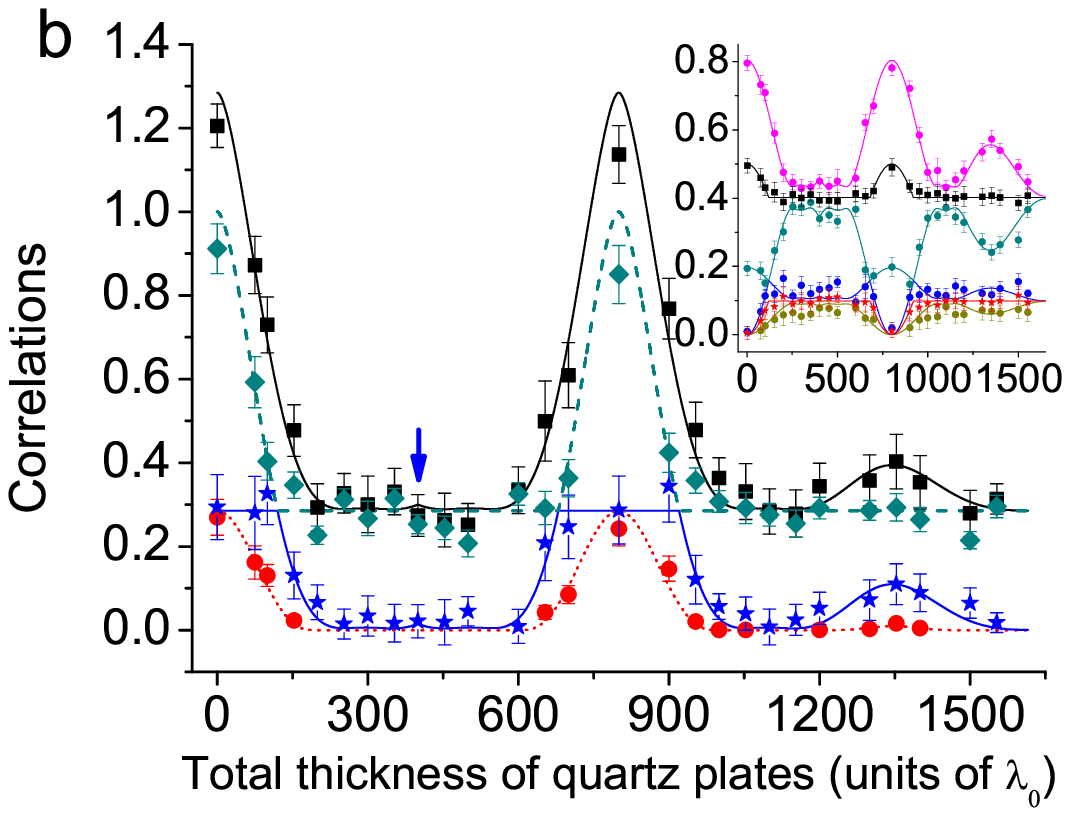}
\end{minipage}
\caption{(Color online). Experimental results of correlation echoes:
({\bf a}) with the $\sigma_{x}$ operation at $200\lambda_{0}$; ({\bf
b}) with the $\sigma_{x}$ operation at $400\lambda_{0}$. Black
squares, dark cyan diamonds, blue stars and red dots represent the
experimental results of $\mathcal{I}$, $\mathcal{C}$, $\mathcal{Q}$
and $REE$ with the black solid line, dark cyan dashed line, blue
solid line and red dotted line representing the corresponding
theoretical predictions, respectively. The $x$ axes represent the
total thickness of Q2 and Q3. The blue arrows identify the
$\sigma_{x}$ operation points. The insets in ({\bf a}) and ({\bf b})
(the $x$ axes represent the total thickness of Q2 and Q3) represent
the corresponding evolution of eigenvalues of $\rho$ and
$\chi_{\rho}$, respectively. The magenta, dark cyan, blue and dark
yellow dots represent the experimental results of $\lambda_{1}$,
$\lambda_{2}$, $\lambda_{3}$ and $\lambda_{4}$ with the magenta,
dark cyan, blue and dark yellow solid lines representing the
corresponding theoretical predictions, respectively. Black squares
and red stars represent the experimental results of
$\frac{1}{2}(\lambda_{1}+\lambda_{2})$ and
$\frac{1}{2}(\lambda_{3}+\lambda_{4})$ with the black and red solid
lines representing the corresponding theoretical predictions,
respectively. $\lambda_{0}=0.78$ $\mu$m.} \label{fig:experiment2}
\end{figure}

It is a great challenge to realize the small FWHM to obtain the
observable revival of classical correlation in the experiment. We
then implement a $\sigma_{x}$ operation on the photon in path $b$ to
investigate the correlation dynamics and get the revival of
classical correlation. The thickness of Q2 is first increased to
$200\lambda_{0}$ or $400\lambda_{0}$ followed by a $\sigma_{x}$
operation which is fulfilled by the HWP2. The thickness of Q3 is
then increased from zero without further increase of Q2 to get the
corresponding dynamics of correlations, which are shown in Fig.
\ref{fig:experiment2}a (Q2=200$\lambda_{0}$) and Fig.
\ref{fig:experiment2}b (Q2=$400 \lambda_{0}$), respectively (the $x$
axes represent the total thickness of Q2 and Q3). The black squares,
dark cyan diamonds, blue stars and red dots represent the
experimental results of $\mathcal{I}$, $\mathcal{C}$, $\mathcal{Q}$
and $REE$ with the black solid line, dark cyan dashed line, blue
solid line and red dotted line representing the corresponding
theoretical predictions, respectively. We can see that correlation
echoes are formed when Q3=$200\lambda_{0}$ in Fig.
\ref{fig:experiment2}a and Q3=$400\lambda_{0}$ in Fig.
\ref{fig:experiment2}b, {\it i. e.}, at the time when Q3 is
increased to the same thickness as Q2. This implies that the
dephasing effect between $|H\rangle$ and $|V\rangle$ caused by Q2 is
completely compensated by Q3 and all the correlations maximally
revives to the initial values. During this process, the sudden
transition from quantum to classical revival area is observed (Fig.
\ref{fig:experiment2}a at the thickness of about $280\lambda_{0}$
and Fig. \ref{fig:experiment2}b at the thickness of about
$680\lambda_{0}$). The classical correlation revives from the
constant period to its maximal value, whereas the quantum
correlation stops revival and remains constant. If we further
increase Q3, the subsequent dynamics of correlations are just the
same as that shown in Fig. \ref{fig:experiment}a. The blue arrows in
the figures identify the $\sigma_{x}$ operation points. The
correlation dynamics between the initial state and the maximally
revived states are symmetric about the $\sigma_{x}$ operation point,
which is similar to the phenomenon of spin echo in nuclear magnetic
resonance \cite{Hahn50}. The insets in Fig. \ref{fig:experiment2}a
and Fig. \ref{fig:experiment2}b represent the corresponding dynamics
of eigenvalues of $\rho$ and $\chi_{\rho}$. The magenta, dark cyan,
blue and dark yellow dots represent the experimental results of
$\lambda_{1}$, $\lambda_{2}$, $\lambda_{3}$ and $\lambda_{4}$ with
the magenta, dark cyan, blue and dark yellow solid lines
representing the corresponding theoretical predictions,
respectively. Whereas the black squares and red stars represent the
experimental results of $\frac{1}{2}(\lambda_{1}+\lambda_{2})$ and
$\frac{1}{2}(\lambda_{3}+\lambda_{4})$ with the black and red solid
lines representing the corresponding theoretical predictions,
respectively. We can see that the eigenvalues also display the echo
effect and the second switch in the second maximal eigenvalue
$\lambda_{2}$ in both the insets of Fig. \ref{fig:experiment2}a and
\ref{fig:experiment2}b leads to the revival of classical
correlation.

\section{Discussion and Conclusion}

The reasons for quantum advantage in quantum information processing
are still controversial \cite{Vedral09}. Recently, a simple
one-to-one relationship between bipartite entanglement of formation
\cite{Bennett96} and quantum discord in a general tripartite system
is proposed \cite{Fanchini102}. By taking the environment which is
initially maximally entangled to the system with the reduced
maximally mixed state in the DQC1 protocol into consideration, it is
suggested that both the quantum discord and entanglement are
responsible for the quantum computer speedup \cite{Fanchini102}.
Inspired by the factorization law of entanglement evolution in noisy
quantum channels \cite{Konrad08,Farias09,Xu09}, it is expected that
similar simple relationship for quantum correlation under noise
environment exists. With the increasing interests, there will be
more distinctive discoveries in this field.

In our experiment, we have investigated the correlation dynamics in
a non-Markovian dephasing environment. The whole point of a
non-Markovian environment is that it retains a memory of a system at
a given time and then later passes this information back into the
system in some form or another. In this experiment, the
non-Markovian environment acts via the FP cavity followed by quartz
plates on the biphoton system of only one of the photons. Due to the
discrete frequency distribution of the photon which leads to the
refocusing of the relative phase in the dephasing environment, the
non-Markovian effect occurs and the revival of quantum correlation
is obtained from near zero area. However, the non-Markovian effect
is too weak to revive the classical correlation. With the narrower
FWHM of the wave packets of the discrete frequency distribution, the
revival of classical correlation would be also achieved. On the
other hand, if the FWHM of the wave packets become lager, the
non-Markovian effect becomes weaker. When the discrete frequency
distribution tends to continuous Gaussian distribution, we would get
the Markovian limited dynamics. We further implement a $\sigma_{x}$
operation on the photon in path $b$ to investigate the corresponding
correlation dynamics. During this process, we obtain correlation
echoes, in which the sudden transition from quantum to classical
revival regime is observed. This work is a useful and informative
add-on to our previous works, in which the entanglement collapse and
revival is observed \cite{Xu102} and the Markovian limited dynamics
of correlations is demonstrated \cite{Xu10}. The method can be used
to control the revival time of correlations, which would find
important applications in quantum memory.

We thank Cheng-Hao Shi for help in performing the experiment.  This
work was supported by National Fundamental Research Program,
National Natural Science Foundation of China (Grant Nos. 60921091,
10734060, 10874162, 11004185), China Postdoctoral Science Foundation
(Grant No. 20100470836) .


\end{document}